\documentclass[journal=nalefd,manuscript=article]{achemso}

\usepackage{achemso}
\usepackage{graphicx}
\usepackage{amsmath}
\usepackage{mathtools}
\usepackage{bbold}
\usepackage{braket}
\usepackage{bm}
	\usepackage{hyperref}		% hyperlinks in the pdf, has to be (almost) last %
	\usepackage[figure]{hypcap}		% corrects hyperlinks to figures %
	\hypersetup{
		unicode		= false,	% non-Latin characters in Acrobat's bookmarks
		pdfborder		= {0 0 0}	% style of the border around a link; {0 0 0} gives no border %
		pdftoolbar		= true,		% show Acrobat's toolbar %
		pdfmenubar		= true,		% show Acrobat's menu %
		pdffitwindow		= false,     	% window fit to page when opened %
		pdfstartview		= {FitH},    	% fits the width of the page to the window %
		pdfnewwindow	= true,      	% links in new window %
		colorlinks		= true,		% false: boxed links; true: colored links %
		linkcolor		= blue,		% color of internal links %
		citecolor		= blue,		% color of links to bibliography %
		filecolor		= blue,		% color of file links %
		urlcolor		= blue,		% color of external links %
		linktocpage		= true,		% whole entry or only page in the toc is made into a link %
		pdftitle		= {?},
		pdfauthor		= {?},
		pdfsubject		= {Raman signatures of electronic excitations in rhombohedral graphite},
		pdfcreator		= {MikTeX},
		pdfproducer		= {Marcin Mucha-Kruczynski},
		pdfkeywords		= {},
}

\usepackage{cleveref}

	\newcommand{\vect}[1]{\boldsymbol{#1}}
	\newcommand{\op}[1]{\hat{\boldsymbol{#1}}}

	\newcommand{\Raman}{\omega}
	\newcommand{\onlinecite}[1]{\hspace{-1 ex} \nocite{#1}\citenum{#1}}
\title{Spectroscopic Signatures of Electronic Excitations in Raman Scattering in Thin Films of Rhombohedral Graphite}
\author{Aitor Garc\'{i}a-Ruiz}
\affiliation{Department of Physics, University of Bath, Claverton Down, BA2~3FL, United Kingdom}
\author{Sergey Slizovskiy}
\affiliation{National Graphene Institute, University of Manchester, Booth St E, Manchester, M13 9PL, United Kingdom}
\alsoaffiliation{Department of Physics and Astronomy, University of Manchester, Oxford Road, Manchester, M13 9PL, United Kingdom}
\author{Marcin Mucha-Kruczy\'{n}ski}
\affiliation{Department of Physics, University of Bath, Claverton Down, BA2~3FL, United Kingdom}
\alsoaffiliation{Centre for Nanoscience and Nanotechnology, University of Bath, Claverton Down, BA2~3FL, United Kingdom}
\email{m.mucha-kruczynski@bath.ac.uk}
\author{Vladimir I.~Fal'ko}
\affiliation{National Graphene Institute, University of Manchester, Booth St E, Manchester, M13 9PL, United Kingdom}
\alsoaffiliation{School of Physics and Astronomy, University of Manchester, Oxford Road, Manchester, M13 9PL, United Kingdom}
\alsoaffiliation{Henry Royce Institute, Manchester, M13 9PL, United Kingdom}

\keywords{rhombohedral graphite, graphene, Raman spectroscopy, electronic Raman scattering, optical absorption, stacking-dependent properties}

\begin{document}

\begin{abstract}
Rhombohedral graphite features peculiar electronic properties, including persistence of low-energy surface bands of a topological nature. Here, we study the contribution of electron-hole excitations towards inelastic light scattering in thin films of rhombohedral graphite. We show that, in contrast to the featureless electron-hole contribution towards Raman spectrum of graphitic films with Bernal stacking, the inelastic light scattering accompanied by electron-hole excitations in crystals with rhombohedral stacking produces distinct features in the Raman signal which can be used both to identify the stacking and to determine the number of layers in the film.
\end{abstract}

\maketitle

\newpage

After seven decades of intense studies \cite{dresselhaus_advphys_2002, kelly_book_1981, haering_canjphys_1958}, graphite still surprises us by the richness of physical phenomena and optoelectronic effects it can host. The agility of graphite is, largely, due to the van der Waals (vdW) nature of its interlayer bonding, coexisting with a substantial hybridisation between the electronic states in the consecutive layers. On the one hand, a weak vdW bonding between honeycomb graphene layers in graphite allows for the formation of various stacking configurations: naturally appearing Bernal, rhombohedral and turbostratic graphites \cite{matuyama_nature_1956, lui_nanolett_2011}, or designer-twisted graphene bilayers \cite{kim_nanolett_2016}. On the other hand, the interlayer hybridization of carbon $P_{z}$ orbitals with a characteristic energy $\sim 0.3$-$0.4$ eV, combined with a peculiar Dirac-like bands of electrons near the Fermi level in graphene \cite{wallace_physrev_1947, mcclure_physrev_1956, novoselov_nature_2005}, makes electronic properties of various graphitic structures distinctively different. As a result, over the years, many extreme regimes of quantum transport and correlations were found in ultrathin graphitic films \cite{novoselov_natphys_2006, ki_nanolett_2014, kim_natphys_2019, grushina_natcomms_2015, nam_science_2018, nam_2dmaterials_2016, yin_natphys_2019, pamuk_prb_2017} and artificially fabricated structures \cite{cao_nature_2018, cao_nature_2018_2, ahn_science_2018, yankowitz_science_2019, kim_pnas_2017}.

\begin{figure}[t]
\includegraphics[width=0.5\columnwidth]{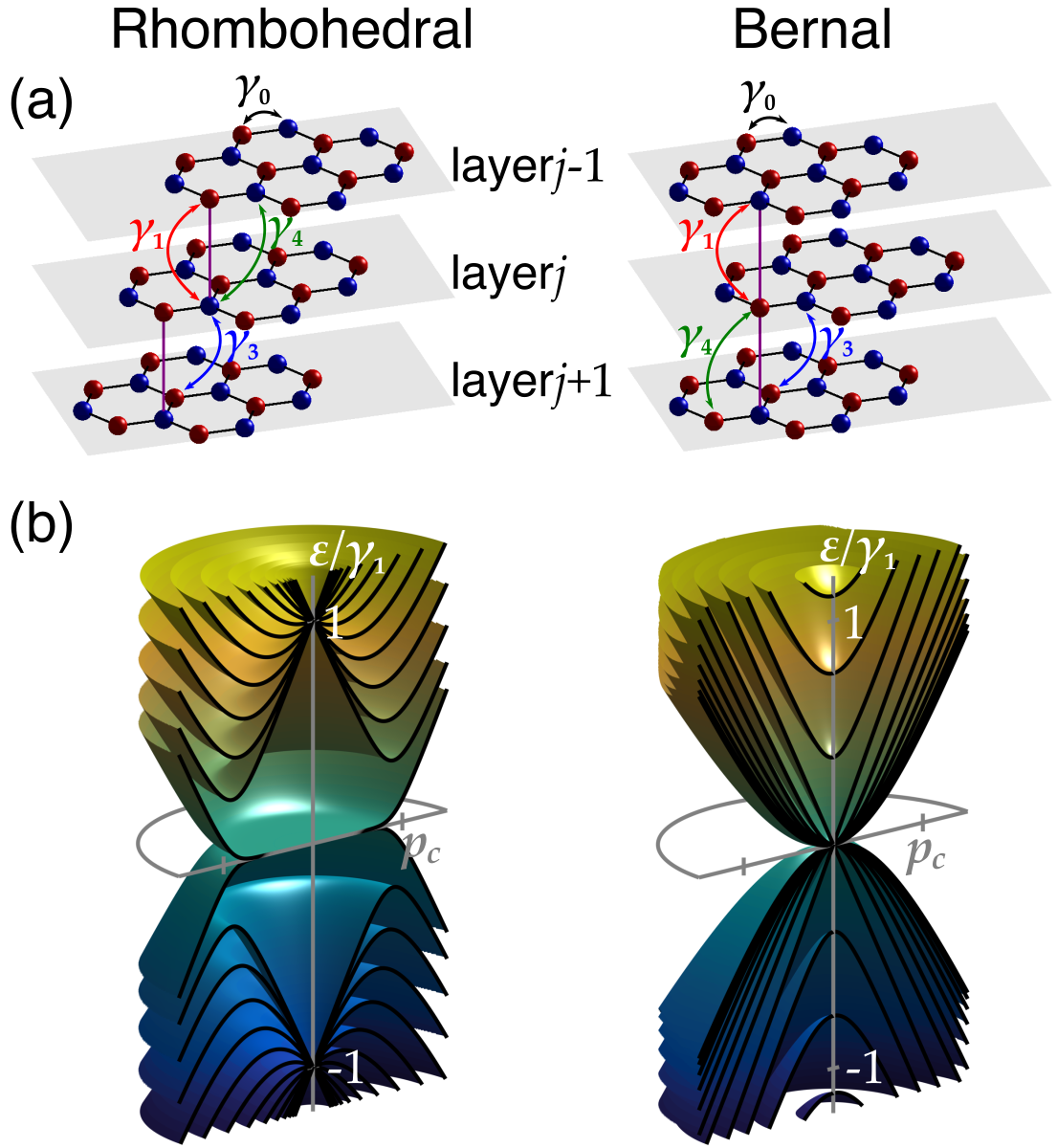}
\caption{Comparative overview of rhombohedral (left) versus Bernal (right) graphite. (a) Lattice structure and relevant hoping couplings. (b) Low-energy subbands in a 15-layer-thick film. For a film of rhombohedral graphite, the subbands feature van Hove singularities which are split from the subband edges for finite $\gamma_{3}$ and $\gamma_{4}$, and the same hoppings make the lowest energy subband dispersion non-monotonic and trigonally warped. Moreover, momentum space positions of van Hove singularities in different subbands (including the lowest energy one) all differ; as a result, spectroscopic features in intersubband absorption and Raman arise from different areas of momentum space.}\label{fig:stacking}
\end{figure}

Rhombohedral is a structural phase of graphite which has a specific 'ABC'  stacking of consecutive honeycomb layers of carbon atoms, such that every atom has a nearest neighbour from an adjacent layer either directly above or underneath it, with which their $P_{z}$ orbitals hybridise with a coupling $\gamma_{1}\approx 0.39$meV \cite{mccann_reps_2013}. This distinguishes it from Bernal's 'ABA' stacking, where half of carbons find the closest neighbours in the consecutive layer, whereas the other half appears to be between the empty centres of the honeycombs in the layers above and below. This difference between the crystalline structures of rhombohedral and Bernal graphite is depicted in Fig.~\ref{fig:stacking}(a), together with the intra- and interlayer hopping couplings between carbon $P_{z}$ orbitals, marked according to the Slonczewski-Weiss-McClure (SWMcC) tight-binding model parametrisation \cite{mcclure_physrev_1957, slonczewski_physrev_1958}. The difference in the lattice structures determines the difference between electron band structures \cite{haering_canjphys_1958, guinea_prb_2006, koshino_prb_2009} of these two phases of graphite, illustrated in Fig.~\ref{fig:stacking}(b) for their 15-layer-thick films. The most pronounced difference in these spectra is related to the existence of low-energy subbands ($0^{+}$ and $0^{-}$), confined to the top/bottom surfaces of a film of ABC graphite \cite{guinea_prb_2006, koshino_prb_2009}, which are almost flat over the momentum range, $\sim p_c = \gamma_1/v$, around $\vect{K}$ and $\vect{K}'$ valleys ($v \approx 10^8$ cm/s is Dirac velocity of electrons in graphene). These low-energy surface bands, already established using angle-resolved photoemission spectroscopy \cite{coletti_prb_2013, pierucci_acsnano_2015, henck_prb_2018}, raise expectations for the formation of strongly correlated (magnetic \cite{pamuk_prb_2017} or superconducting \cite{kopnin_prb_2013}) states, reigniting the interest in ABC graphitic films \cite{sugawara_npgam_2018, wang_nanolett_2018, myhro_2dmaterials_2018, chen_natphys_2019, yin_prl_2019}.

Raman spectroscopy is, currently, one of the methods of choice used for the identification of structural properties of atomically thin films of van der Waals materials \cite{ferrari_natnano_2013, zhang_csrevs_2018, bhimanapaty_acsnano_2015}, providing information about the number of layers in the film, strain and doping.  Usually, such Raman spectroscopy detects phonon excitations in the lattice, which also stands for the ABC graphitic films \cite{henni_nanolett_2016, torche_prmat_2017}, but with a lesser clarity of interpreting the data as compared to Bernal graphite \cite{latychevskaia_fop_2019}. At the same time, it has been demonstrated that Raman scattering enables one to access directly interband electronic excitations in monolayer \cite{kashuba_prb_2009, faugeras_prl_2011, kuhne_prb_2012} and bilayer \cite{mucha-kruczynski_prb_2010, riccardi_prmat_2019} graphene, though application of a strong magnetic field leading to Landau level quantization was required to highlight the spectral features of electron-hole excitations. Here, we show that the peculiar dispersion of electrons in thin films of rhombohedral graphite produces peaks in their Raman response at energies,
\begin{align}\label{eqn:transitions}
\omega_{n}\approx 4\gamma_{1}\sin\frac{(n+\frac{1}{2})\pi}{2N+1},\,\, 1\leq n\leq \left\lfloor \frac{N}{2}\right\rfloor,
\end{align}
where $\lfloor x\rfloor$ is the greatest integer less than or equal to $x$. These features, which are related to van Hove singularities in the vicinity of subband edges in the thin film spectrum, can be used for the identification of the number of layers, $N$, in the ABC graphitic films. In addition, we find that Raman can detect presence of stacking faults in thin films of rhombohedral graphite.

To model optical properties of graphitic films, we use a brute-force diagonalization of a hybrid 'k$\cdot$p' tight-binding model (HkpTB) in which the intra-layer hopping of electrons between carbon atoms is taken into account in a continuous description of sublattice Bloch states using 'k$\cdot$p' theory near the $\vect{K}$ and $\vect{K}'$ valleys, combined with interlayer hopping introduced in the spirit of a tight binding model. The Hamiltonian, written in the basis $\{\phi_{A1\xi},\phi_{B1\xi},\phi_{A2\xi},\phi_{B2\xi},...,\phi_{AN\xi},\phi_{BN\xi}\}$ of Bloch states $\phi_{Xj\xi}$, constructed of $P_{z}$ orbitals on $A$ and $B$ sublattice of honeycomb lattice of the $j$-th graphene layer, reads
\begin{align}\label{eqn:hamiltonian}
& \op{H}=\op{H}_{0}+\op{H}_{T},\\
& \op{H}_{0}=v\op{I}_{N}\otimes\begin{pmatrix} 0 & \op{\kappa}^{\dagger} \\ \op{\kappa} & 0\end{pmatrix},\,\,\op{\kappa}=\xi p_{x}+ip_{y},\nonumber\\
& \op{H}_{T} = \left(
\begin{matrix}
0 & \op{T}_{1,2} & 0 & \dots & 0 & 0 \\
{\op{T}}_{1,2}^{\dagger} & 0 & \op{T}_{2,3} & \dots & 0 & 0\\
0 & \op{T}_{2,3}^{\dagger} & 0 & \dots & 0 & 0\\
\vdots & \vdots & \vdots & \ddots & \vdots & \vdots\\
0 & 0 & 0 & \dots & 0 & \op{T}_{N-1,N} \\
0 & 0 & 0 & \dots & \op{T}^{\dagger}_{N-1,N} & 0
\end{matrix}\right)\nonumber.
\end{align}
Here, $\vect{\sigma}=(\sigma_{x},\sigma_{y},\sigma_{z})$ is the vector of Pauli matrices, $\op{I}_{N}$ is the $N\times N$ identity matrix, $\xi=\pm 1$ denotes the two inequivalent valleys $\vect{K}=(\tfrac{4\pi}{3a},0)$ and $\vect{K}'=(-\tfrac{4\pi}{3a},0)$ ($a=2.46$ {\AA} is the graphene lattice constant), and $\vect{p}=(p_{x},p_{y})$ is the electron momentum in the valley. The $2\times 2$ matrices $\op{T}_{j,j+1}$ describe the electron hopping between consecutive layers. For the rhombohedral stacking, 
\begin{align*}
\op{T}_{j,j+1}=\op{T}\equiv
\left(\begin{matrix}
-v_{4}\op{\kappa}^{\dagger} & v_{3} \op{\kappa} \\
\gamma_1 & -v_{4} \op{\kappa}^{\dagger}
\end{matrix}\right),
\end{align*}
$v_{3(4)}=\tfrac{a\sqrt{3}}{2\hbar}\gamma_{3(4)}$, $\gamma_{3}\approx0.26$ eV and $\gamma_{4}\approx0.2$ eV, whereas for Bernal stacking, 
\begin{align*}
\op{T}_{j,j+1}=\left\{ \begin{matrix} \op{T}\,\,\mathrm{for}\,\,\mathrm{odd}\,\,j, \\ \op{T}^{\dagger}\,\,\mathrm{for}\,\,\mathrm{even}\,\,j. \end{matrix} \right.
\end{align*}

\begin{figure}[t]
\includegraphics[width=0.5\columnwidth]{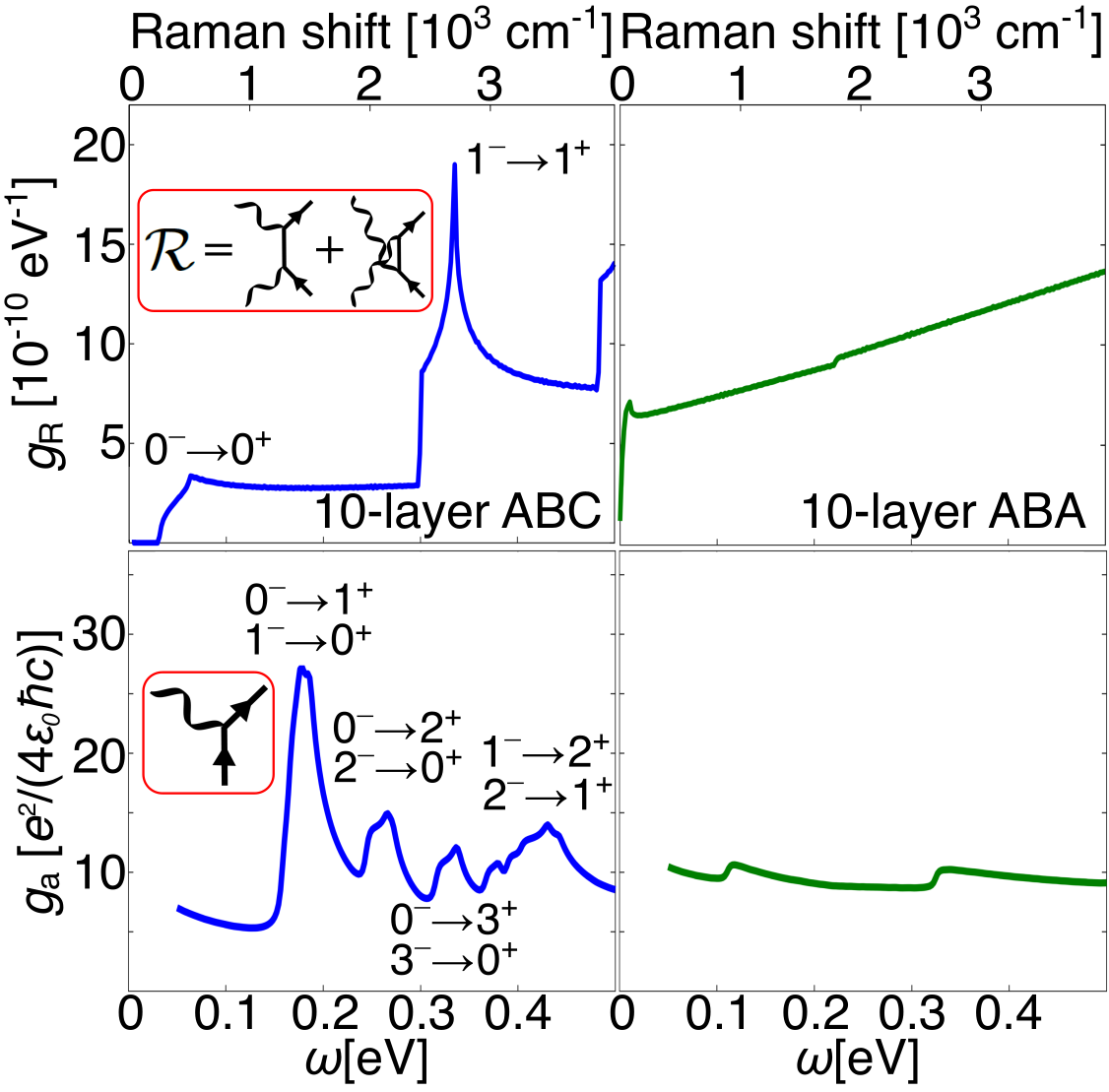}
\caption{Spectral density of electronic Raman scattering $g_{\mathrm{R}}(\Raman)$ (for excitation with photons with $\Omega=2$ eV) and absorption coefficient $g_{\mathrm{a}}(\omega)$ of 10-layer rhombohedral (ABC, left panel) and Bernal (ABA, right panel) graphite. For absorption, we assumed level broadening of $3$ meV and plot the spectrum from $\omega>50$ meV. The spectral density of Raman reflects the joint density of states of electrons and holes. It features distinct peaks that correspond to van Hove singularities in the subbands preceded by steps that arise from the subband edge. In the models that ignore trigonal warping, those two features would overlay, but for finite $\gamma_{3}$ and $\gamma_{4}$ the subband van Hove singularities in conduction/valence lie above/below the corresponding subband edges.}\label{fig:10layers}
\end{figure}

The absorption of incident light, arriving perpendicular to the film and characterised by vector potential $\vect{A}_{\omega}=\vect{E}/i\omega$ with in-plane polarization $\vect{l}=(l_{x},l_{y})$, by a thin graphitic film (undoped and with the thickness less than attenuation length) is described by absorption coefficient \cite{abergel_apl_2007, abergel_prb_2007},
\begin{align}\label{eqn:absorption}
g_{\mathrm{a}}(\omega)\approx\frac{8\alpha}{\pi\omega}\Im\sum_{n^{-},m^{+}}\int\mathrm{d}\vect{p}\frac{\lvert\braket{\vect{p},m^{+}|\frac{\partial\op{H}_{0}}{\partial\vect{p}}\cdot\vect{l}|\vect{p},n^{-}}\rvert^{2}}{\omega+\epsilon_{\vect{p},n^{-}}-\epsilon_{\vect{p},m^{+}}+i0}.
\end{align}
Here, $\ket{\vect{p},n^{s}}$ are states (with momentum $\vect{p}$) in the $n$-th subband on the conduction/valence ($s=\pm$) band side at energy $\epsilon_{\vect{p},n^{s}}$ and $\alpha=\tfrac{e^{2}}{4\pi\epsilon_{0}\hbar c}=\tfrac{1}{137}$ is the fine structure constant. By inspection of the matrix structure of the operator $\tfrac{\partial\op{H}_{0}}{\partial\vect{p}}$ and the eigenstates of the Hamiltonian in Eq.~\eqref{eqn:hamiltonian}, we find that (similarly to bilayer graphene \cite{abergel_prb_2007}), the dominant valence-conduction band transition are such that $0^{-}(n^{-})\rightarrow n^{+}(0^{+})$ and $n^{-} \rightarrow (n+1)^{+}$ and $n^{-} \rightarrow (n-1)^{+}$, resulting in distinct features at $\omega_{n}\approx2\gamma_{1}\sin\tfrac{(n+\tfrac{1}{2})\pi}{2N+1}$ and $\omega'_{n}\approx 4\gamma_{1}\sin\tfrac{(n+1)\pi}{2N+1}$, respectively, with $0\leq n\leq \left\lfloor \frac{N}{2}\right\rfloor$ (for $N\gg 1$), marked in Fig.~\ref{fig:10layers}. To mention, in Ref.~\onlinecite{li_prl_2012}, some IR absorption features have been observed in rhombohedral graphite flakes identified as with 4, 5 and 6 layers that were interpreted as the $0^{-}(1^{-}) \rightarrow 1^{+}(0^{+})$ sequence. While for small $N$ formulae for $\omega_{n}$ and $\omega'_{n}$ overestimate the peak positions, computing absorption spectra numerically would reproduce the measured spectra once we take $\gamma_{1}=0.32$ eV, which is less than the values 0.38--0.4 eV typically quoted for bilayer graphene and in Slonczewski-Weiss-McClure Bernal graphite \cite{mccann_reps_2013}.  

In inelastic scattering, a photon with energy $\Omega$, arriving to the sample at normal incidence, scatters to a photon with energy $\Omega'=\Omega-\omega$, leaving behind an electron-hole excitation with energy $\omega$. Specifically for graphite, the amplitude of this process is dominated by the sum of two amplitudes presented in the form of Feynman diagrams in the inset in Fig.~\ref{fig:10layers}(a), which would cancel each other (due to the opposite sign, $\Omega$ and $-\Omega^{\prime}$, of the energy mismatch in the intermediate state) for non-relativistic electron in a simple metal with a parabolic dispersion, but, for Dirac electrons, generate an amplitude \cite{kashuba_prb_2009, mucha-kruczynski_prb_2010}, 
\begin{equation}
\mathcal{R}\approx i\frac{(e\hbar v)^2}{2\epsilon_0\Omega^2}(\vect{l}\times\vect{l}')_z \op{I}_{N}\otimes\sigma_z.
\end{equation} 
The latter expression means that the main contribution to Raman comes from $n^{-} \rightarrow n^{+}$ intersubband transitions and indicates that, in the measurements, the inelastic light scattering leaving behind electron-hole excitations can be filtered out by picking up the cross-polarisation component of the Raman signal: the outgoing photon would be linearly polarized in the direction perpendicular to the linear polarization of the incoming photon. Then, spectral density of Raman scattering in a film with the Fermi level at the edge between the $n=0^{\pm}$ subbands is
\begin{align}\label{eqn:ers}
g_{\mathrm{R}}(\Raman) & =\frac{1}{c}\int\frac{\mathrm{d}\vect{q}'\,w(\Raman)}{(2\pi\hbar)^3}\delta(\Omega'-c\vect{q}')=\frac{\Omega^{2}}{(2\pi\hbar)^{3} c^4}w(\Raman),\\
w(\Raman) & =\frac{2}{\pi\hbar^{3}}\sum_{sn,s'm}\int\mathrm{d}\vect{p}\left|\Bra{\vect{p},m^{+}}\mathcal{R}\Ket{\vect{p},n^{-}}\right|^2\times\delta(\epsilon_{\vect{p},m^{+}}-\epsilon_{\vect{p},n^{-}}-\Raman),\nonumber
\end{align}
with an overall quantum efficiency $I=\int \mathrm{d}\Raman g_{\mathrm{R}}(\Raman)\sim 10^{-10}$, which was proven to be in the measurable range by the earlier studies of graphene \cite{kashuba_prb_2009, faugeras_prl_2011, kuhne_prb_2012, mucha-kruczynski_prb_2010, riccardi_prl_2016, riccardi_prmat_2019}.

\begin{figure}[t]
\includegraphics[width=1.0\columnwidth]{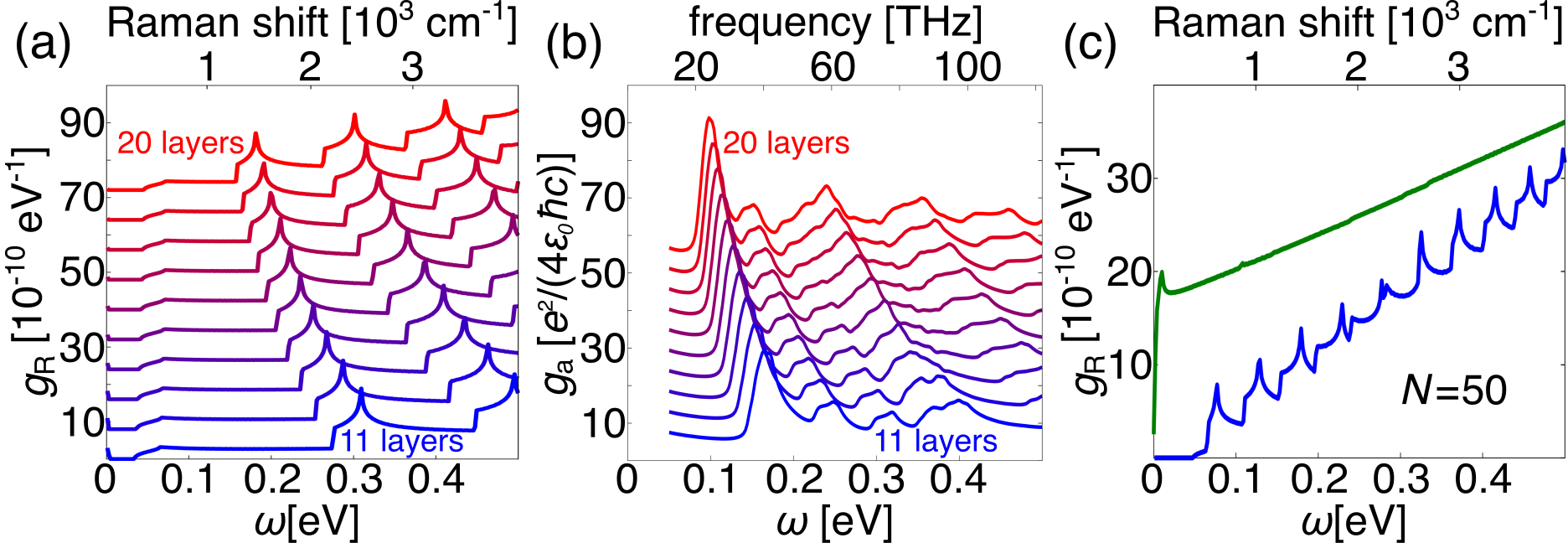}
\caption{Raman signature of rhombohedral graphite. (a) Spectral density, $g_{\mathrm{R}}(\Raman)$, of electron excitations in Raman scattering of photons with $\Omega=2$ eV as a function of film thickness. (b) THz absorption, $g_{\mathrm{a}}(\omega)$, for the same films. The consecutive curves are shifted up by $8\times 10^{-10}$ eV$^{-1}$ in (a) and by $5e^{2}/(4\epsilon_{0}\hbar c)$ in (b). (c) Electronic Raman scattering on 50-layer-thick rhombohedral (blue) and Bernal (green) graphite films.}\label{fig:thicknessevolution}
\end{figure}

Figure \ref{fig:10layers} exemplifies the calculated Raman scattering and THz/IR absorption spectra in films of rhombohedral and Bernal graphite. While their spectra are essentially featureless for Bernal stacking, for ABC stacking they show a series of peaks related to the excitations of electrons between van Hove singularities in (i) the $n$-th valence band to the $n$-th conduction band transition for Raman and (ii) the $n$-th valence band to the $(n\pm 1)$-th conduction band transition for absorption. Figure \ref{fig:thicknessevolution} shows the Raman and THz absorption spectra for the films of rhombohedral graphite with various thicknesses, where the positions of Raman peaks coincide with the values described by Eq.~\eqref{eqn:transitions}. A Raman feature, interpreted as an intersubband excitation in $N=15$--$17$ layers rhombohedral graphite film has been reported in Ref.~\onlinecite{henni_nanolett_2016}; when compared to the results in Fig.~\ref{fig:thicknessevolution}, the measured feature (Raman shift 1800--2000 cm$^{-1}$) agrees well with the calculated range 1700--1900 cm$^{-1}$ for the position of van Hove singularity (estimated for $\gamma_{1}=0.39$ eV). Therefore, we suggest that the studies of electronic excitations in Raman scattering (which can be identified by means of cross-polarisation measurement) can be used to distinguish graphitic films with ABC stacking from Bernal graphite, and even to determine their thickness. To mention, for films thicker than those described in Fig.~\ref{fig:thicknessevolution}(a), the attenuation of the photon field inside the film, described by absorption coefficient $g_{\mathrm{a}}(\Omega)\approx\alpha\pi\approx 2.3\%$, would require to take into account the inhomogeneity of the excitation field profile and the outgoing photon field distribution in the calculation of the matrix elements of the Raman process. For the experiments where Raman signal would be detected in the transmission geometry, this would simply lead to the damping of the overall Raman spectrum by the factor of $\exp(-0.046N)$, whereas, for the detection of Raman signal in the reflected light, the ABC film spectrum would additionally change, as shown in Fig.~\ref{fig:thicknessevolution}(c) for a 50-layer thick films, losing the distinguishing features of the rhombohedral stacking. Additionally, we note that disorder and a finite scattering rate $\tau^{-1}$ for electrons would broaden the spectral features in Raman to a $\hbar\tau^{-1}$ linewidth. 

\begin{figure}[t]
\includegraphics[width=0.5\columnwidth]{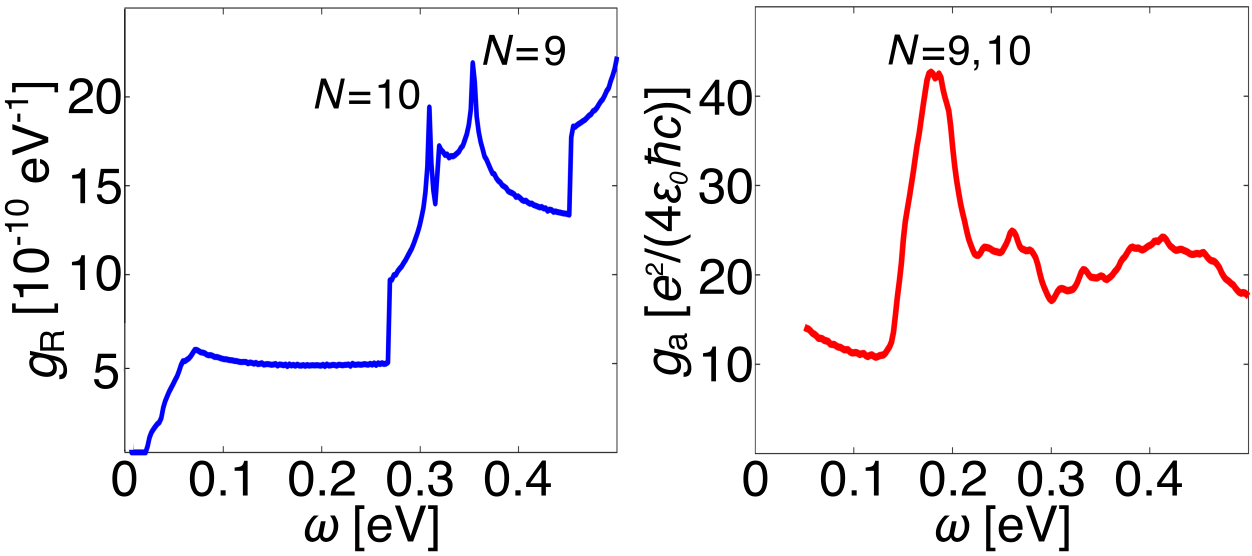}
\caption{Rhombohedral graphitic film with a stacking fault. Spectral density of Raman, $g_{\mathrm{R}}(\Raman)$, and infra-red absorption coefficient, $g_{\mathrm{a}}(\omega)$, of a 20-layer-thick rhombohedral film with a fault at layers 10/11.}\label{fig:fault}
\end{figure}

Finally, in anticipation of possible stacking faults in rhombohedral graphite, we computed Raman and absorption spectra of ABC films with an ABA stacking fault in the middle of it. Introduction of a Bernal stacking between the $j$-th and $(j+1)$-th layers of a film is taken into account by a change of one of the hopping matrices, $\op{T}_{j,j+1}$, in Eq.~\eqref{eqn:hamiltonian}, from $\op{T}$ to $\op{T}^{\dagger}$. The resulting spectra, illustrated in Fig.~\ref{fig:fault} for $N=20$ with a fault between the 10th and 11th layers show that the film appears in Raman as overlaying rhombohedral crystals of thicknesses $j$ and $(N-j-1)$ layers. Similarly, an $M$-layer 'ABC insert' in a thin film of Bernal graphite would produce features in the overall Raman spectrum of the film, with the van Hove singularities peaks typical for the $M$-layer film of rhombohedral phase superimposed over the featureless background of Bernal graphite spectrum.

\section{\label{sec:level5}Acknowledgments}
This work has been supported by the UK Engineering and Physical Sciences Research Council (EPSRC) through the Centre for Doctoral Training in Condensed Matter Physics (CDT-CMP), Grant No. EP/L015544/1, as well as EPSRC Grants EP/S019367/1, EP/P026850/1 and EP/N010345/1, the European Graphene Flagship project, European Research Council Synergy Grant {\it Hetero2D}, the Royal Society and Lloyd's Register Foundation Nanotechnology Programme.

%\newpage
%
%\begin{center} 
%\includegraphics{FigToC.pdf}
%\end{center}

\end{document}